\documentclass[conference]{IEEEtran}

\usepackage{graphicx}
\usepackage{algorithmic}
\usepackage{algorithm}

\ifCLASSINFOpdf

\else

\fi

\begin{document}
%
\title{Checking Behavioral Consistency Constraints for Pervasive Context in Asynchronous Environments}


\author{\IEEEauthorblockN{Yu Huang$^{1,2}$, Jianping Yu$^{1,2}$, Jiannong Cao$^3$, Xiaoxing Ma$^{1,2}$, Xianping Tao$^{1,2}$, Jian Lu$^{1,2}$}
\IEEEauthorblockA{$^1$State Key Laboratory for Novel Software Technology\\
Nanjing University, Nanjing 210093, China\\
$^2$Department of Computer Science and Technology\\
Nanjing University, Nanjing 210093, China\\
\{yuhuang, xxm, txp, lj\}@nju.edu.cn, yujianping@ics.nju.edu.cn}
$^3$Internet and Mobile Computing Lab, Department of Computing\\
Hong Kong Polytechnic University, Hong Kong, China\\
csjcao@comp.polyu.edu.hk
}

\maketitle

\begin{abstract}

Context consistency checking, the checking of specified constraint on properties of contexts, is essential to context-aware applications. In order to delineate and adapt to dynamic changes in the pervasive computing environment, context-aware applications often need to specify and check behavioral consistency constraints over the contexts. This problem is challenging mainly due to the distributed and asynchronous nature of pervasive computing environments. Specifically, the critical issue in checking behavioral constraints is the temporal ordering of contextual activities. The contextual activities usually involve multiple context collecting devices, which are fully-decentralized and interact in an asynchronous manner. However, existing context consistency checking schemes do not work in asynchronous environments, since they implicitly assume the availability of a global clock or relay on synchronized interactions.

To this end, we propose the Ordering Global Activity (OGA) algorithm, which detects the ordering of the global activities based on predicate detection in asynchronous environments. The essence of our approach is the message causality and its on-the-fly coding as logic vector clocks in asynchronous environments. We implement the Middleware Infrastructure for Predicate detection in Asynchronous environments (MIPA), over which the OGA algorithm is implemented and evaluated. The evaluation results show the impact of asynchrony on the checking of behavioral consistency constraints, which justifies the primary motivation of our work. They also show that OGA can achieve accurate checking of behavioral consistency constraints in dynamic pervasive computing environments.

\end{abstract}

\IEEEpeerreviewmaketitle

\section{Introduction}

Pervasive applications are typically context-aware, using various kinds of contexts, such as location and time, to provide smart services \cite{Romer04, Ran04, Loke09, Mamei09}. Context-aware applications need to monitor whether contexts bear specified property, thus being able to adapt to the pervasive computing environment accordingly \cite{Bellavista03, Capra03, Henricksen04, Hu08}. This brings the essential issue of context consistency checking, i.e. checking of specified constraints on properties of contexts \cite{Xu06, Huang09}.

Context consistency checking has been widely studied in pervasive computing and software engineering communities \cite{Xu05, Xu06, Xu08, Bu06, Bu06a, Huang08, Huang09}. For example in \cite{Xu06}, consistency constraints are expressed in first order logic, and contextual properties like ``location of the user is the meetingroom and a presentation is going on" can be specified. However, existing schemes mainly focus on checking of {\it static} consistency constraints, i.e. constraints delineating properties of contexts at given snapshot of time. Though static consistency constraints can capture interesting properties of the pervasive computing environment, they inherently lack the notions of relative temporal order \cite{Babaouglu95, Kaveti09}. Such constraints cannot characterize {\it behavioral patterns} of contexts, such as ``$C_1$: the user in in his office (detected by the RFID reader and the light sensor in the office); then the user leaves the office (detected by the RFID reader and the light sensor in the corridor)".

The discussions above necessitate the checking of {\it behavioral} consistency constraints, i.e. constraints delineating behavior patters of contexts. The key issue in checking behavioral consistency constraints is how to decide the temporal order among contextual activities. This issue is challenging in pervasive computing environments, mainly due to the following two observations:
\begin{itemize}
  \item Contextual activities are often {\it global}, involving multiple decentralized context collection devices. For example, in constraint $C_1$ discussed above, the location context is decided by two different sensors (RFID reader and light sensor), in order to improve the accuracy of context. Pervasive applications and context collecting devices usually coordinate in a fully-distributed manner, based on wired/wireless communications.
  \item The pervasive computing environment is often {\it asynchronous} \cite{Sama08, Huang09, Kaveti09}. Specifically, context collecting devices do not necessarily have a global clock. They heavily rely on wireless communications, which suffer from uncertain delay. Moreover, due to resource constraints, context collection devices, e.g. battery-powered sensors, often need to buffer context data for certain time. Periodic or adaptive schemes are employed to schedule the dissemination of context data \cite{Sama08}. The different context update rates also result in the asynchrony of pervasive computing environments, which cannot be easily synchronized by message exchanging. However, existing consistency checking schemes implicitly assume that the contexts being checked belong to the same snapshot of time \cite{Xu05, Xu06, Xu08, Bu06, Bu06a}. This assumption does not hold in asynchronous pervasive computing environments.
\end{itemize}

To address the challenges discussed above, we study in this paper the checking of behavioral consistency constraints in asynchronous pervasive computing environments. Specifically,
\begin{itemize}
  \item We define behavioral consistency constraints based on the ordering of global activities. We first define global activities based on the concurrency among local contextual activities on decentralized context collection devices. Then, both the concurrency among local activities and the relative order among global activities are defined based on the happen-before relationship resulting from the message causality in asynchronous environments \cite{Lamport78}.
  \item We propose the {\it Ordering Global Activities} (OGA) algorithm to detect the ordering of global contextual activities and check behavioral consistency constraints. OGA assumes the availability of an underlying middleware infrastructure for asynchronous consistency checking of pervasive context. We have developed such a middleware named {\it Middleware Infrastructure for Predicate detection in Asynchronous environments} (MIPA) \cite{MIPA}, on which OGA can be implemented, deployed and evaluated.
  \item We evaluate OGA in a smart-lock scenario, which is fist investigated in our previous work \cite{Wu08}. The evaluation results show how the asynchrony in the pervasive computing environment affects the checking of behavioral consistency constraints. The results also show the accuracy of OGA in context consistency checking in pervasive computing environments.
\end{itemize}

The rest of this paper is organized as follows. In Section II, we describe our system model. In Section III, we present design of the OGA algorithm. In Section IV and V, we overview the design of MIPA and present the experimental evaluation. Section VI overviews the existing work. In Section VII, we conclude the paper with a brief summary and the future work.

\section{System Model}

In this section, we fist discuss how we model asynchronous pervasive computing environments. Then we discuss how to specify behavioral consistency constraints, which includes specification of global activities and specification of the relative order among global activities. Notations used in the system model are listed in Table \ref{T:Notations-Model}.

\begin{table}
\caption{Notations in the system model}
\label{T:Notations-Model}
\centering
\begin{tabular}{r| p{2.4in}}\hline

Notation & Explanation \\ \hline \hline

$n$ & number of all non-checker processes \\
$m$ & number of global activities \\

$GA_k$ & the $k^{th}$ global activity ($1\leq k \leq m$), which might be either $GA^{AND}_k$ or $GA^{OR}_k$\\
$size(GA_k)$ & number of non-checker processes involved in $GA_k$,  $\sum_{k=1}^{m}size(GA_i) = n$ \\

$P_i$ & the $i^{th}$ non-checker process, $1\leq i \leq n$\\
$P^{(k, j)}$ & the $j^{th}$ non-checker process in $GA_k$, $1\leq j \leq size(GA_k)$ ($P^{(k, j)}$ and $P_i$ are different notations of the same non-checker process) \\

$VC_i$ & vector clock timestamp on $P_i$ \\

$LA_i$ & the $i^{th}$ local activity \\
$LA^{(k,j)}$ & the $j^{th}$ local activity involved in $GA_k$ on $P^{(k,j)}$ ($LA_i$ and $LA^{(k,j)}$ are different notations of the same local activity) \\

$I(GA), I(LA)$ & interval of a global / local activity \\

\hline

\end{tabular}
\end{table}

\subsection{Asynchronous Pervasive Computing Environments}

We model context-aware applications in asynchronous pervasive computing environments as a loosely coupled message-passing system, without any global clock or shared memory. Communications suffer from uncertain delay. Dissemination of context data may be postponed due to resource constraints. We assume that no messages are lost, altered, or spuriously introduced. We do not assume that the underlying communication channel is fist-in-first-out (FIFO). Justifications for the assumptions are discussed in Section III.D.

A context-aware application consists of a collection of processes, among which $n$ {\it non-checker processes} (denoted by $P_1, P_2, \cdots, P_n$) are involved in the checking of behavioral consistency constraints. One {\it checker process} (denoted by $P_{che}$) is dedicated for the checking of context consistency. The consistency checking is based on the classical Lamport's definition of the {\it happen-before} (denoted by `$\rightarrow$') relationship resulting from message causality \cite{Lamport78} and its ``on the fly" coding given by Mattern and Fidge's vector clocks \cite{Mattern89, Fidge88}. Each non-checker process $P_j$ keeps $VC_j$, its own vector clock timestamps. $VC_j[i] (i \neq j)$ is ID of the last message from $P_i$, which has a causal relationship to $P_j$. $VC_j[j]$ for $P_j$ is the next message ID $P_j$ will use. Messages passed in the system can be classified into two types:
\begin{itemize}
  \item {\it Control message}. Non-checker processes send control messages among each other to establish the happen-before relationship among contextual activities.
  \item {\it Checking message}. Non-checker processes send vector clock timestamps of contextual activities via checking messages to the checker process. The checker process decides whether the consistency constraint is satisfied based on the collected timestamps.
\end{itemize}

\subsection{Global Activities}

Contextual activities can be either {\it local} or {\it global}. A local activity takes place on some $P_i$ without any interaction with other processes. We delineate local activities of our concern on non-checker process $P_i$ with local predicate $LA_i$. $LA_i$ is true if the local activity is taking place on $P_i$. Otherwise, it is false. We record the interval in which $LA_i = true$. The false-to-true and the true-to-false transitions (denoted by $\uparrow$ and $\downarrow$ respectively) of $LA_i$ correspond to the beginning and ending of the interval, which are denoted by $I_i.lo$ and $I_i.hi$ respectively.

A global activity results from the interaction among local activities. The interaction projected on the time axis is the concurrency among local activities, i.e., the overlapping of intervals of local activities. To detect whether $I_1, I_2, \cdots , I_n$ overlap, we need to check whether the following Formula (\ref{E:OverlappingIntervals}) is satisfied:
\begin{equation}\label{E:OverlappingIntervals}
  (I_j.lo\rightarrow I_k.hi)\wedge(I_k.lo\rightarrow I_j.hi), \forall 1\leq j\neq k\leq n
\end{equation}

\begin{figure}
\begin{center}
  \includegraphics[width=3in]{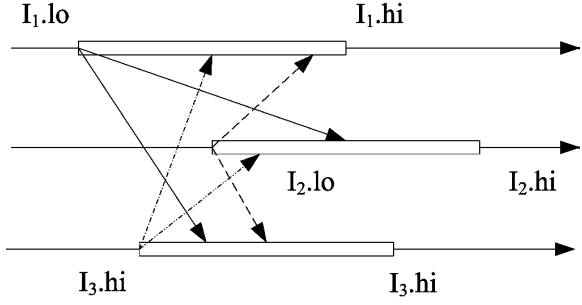}
  \caption{Concurrent local activities}
  \label{F:Overlapping-Intervals}
\end{center}
\end{figure}

\noindent The case of three concurrent local activities is shown in Fig. \ref{F:Overlapping-Intervals}. Detection of concurrent activities has been studied in \cite{Garg96}, as well as in our previous work \cite{Huang09}.

We can further classify global activities based on how we care about the time scope of the interaction, i.e., how we define the interval of the global activity. Specifically, we can define two types of global activities, which are discussed in detail below.
\begin{eqnarray}\label{E:Logic-Definition}
  GA_k &:=& GA^{AND}_k\ |\ GA^{OR}_k \nonumber \\
  GA^{AND}_k &:=& LA^{(k,1)} \wedge \cdots \wedge LA^{(k, size(GA^{AND}_k))} \nonumber \\
  GA^{OR}_k &:=& LA^{(k,1)} \vee \cdots \vee LA^{(k, size(GA^{OR}_k))} \nonumber
\end{eqnarray}

\subsubsection{And-activity}

An and-activity takes place in the period in which multiple local activities are interacting with each other. For example, ``Alice and Bob are in the meeting room" is an and-activity. It takes place in the period when Alice and Bob are both in the meeting room. The interval of an and-activity is defined as the intersection among the intervals of overlapping local activities. For and-activity $GA^{AND}_k = LA^{(k,1)} \wedge LA^{(k,2)} \wedge \cdots \wedge LA^{(k,size(GA_k))}$, its interval is: $$I(GA^{AND}_k) = \bigcap_{1\leq i \leq size(GA_k)}I(LA_i)$$

\begin{figure}
\begin{center}
  \includegraphics[width=3in]{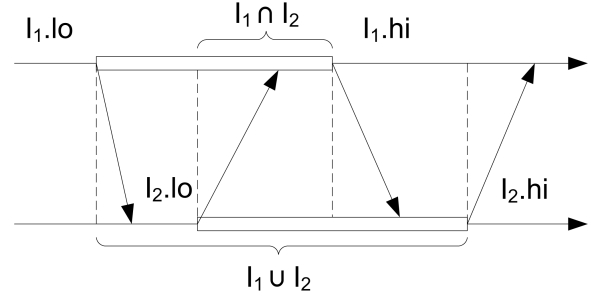}
  \caption{Intervals for and- and or-activities}
  \label{F:Interval-Operations}
\end{center}
\end{figure}

\noindent For example in Fig. \ref{F:Interval-Operations}, $GA = LA_1 \wedge LA_2$. Based on the happen-before relationship established, we have that: $$I(GA) = I_1 \cap I_2 = [I_2.lo, \ I_1.hi]$$

\subsubsection{Or-activity}

An or-activity takes place in the whole period of interaction, i.e., from the happening of the first local activity to the ending of the last local activity. For example, imagine that Alice first waits for Bob in the meeting room. When Bob comes, they have discussions. Then Alice leaves the meeting room. In this case, the or-activity ``Alice or Bob is in the meeting room" takes place in the period starting from the time Alice enters the meeting room and ending at the time Bob leaves. The interval of an or-activity is defined as union of the intervals of overlapping local activities. For or-activity $GA^{OR}_k = LA^{(k,1)} \vee LA^{(k,2)} \vee \cdots \vee LA^{(k,size(GA_k))}$, its interval is defined as: $$I(GA^{OR}_k) = \bigcup_{1\leq i \leq size(GA_k)}I(LA_i)$$

\noindent For example in Fig. \ref{F:Interval-Operations}, if we define $GA' = LA_1 \vee LA_2$, we have that: $$I(GA') = I_1 \cup I_2 = [I_1.lo, \ I_2.hi]$$

\subsection{Ordering Global Activities}

Due to the distributed nature of contexts, we often rely on global activities to delineate the static properties of contexts. To delineate the behavioral patterns of contexts, applications are interested in (global) activities which take place in specified temporal order, such as ``$GA_1$ happens, then $GA_2$ happens, ..., finally $GA_m$ happens". For example, in the behavioral consistency constraint $C_1$ discussed in Section I, the application is interested in the relative order between two global activities ``the user is in the office" and ``the user is in the corridor (leaves the office)". A sequence of ordered global activities is defined as:
\begin{eqnarray}\label{E:Logic-Definition}
  S_{GA} &:=& GA_1 \prec GA_2 \prec \cdots \prec GA_m \nonumber
\end{eqnarray}

\noindent Here, $GA_k$ proceeds $GA_{k+1}$ is defined as the happen-before relationship between the corresponding intervals:
\begin{eqnarray}\label{E:Ordering-2GA}
  GA_k \prec GA_{k+1} := I(GA_k).hi \rightarrow I(GA_{k+1}).lo
\end{eqnarray}

In the next section, we discuss how to check the ordering of global activities in asynchronous pervasive computing environments.

\section{Ordering Global Activities in Asynchronous Pervasive Computing Environments}

In this section, we present design of the proposed {\it Ordering Global Activities} (OGA) algorithm. The OGA algorithm consists of three parts: 1) the non-checker process specifies the message activities upon changes in the local predicate value; 2) the checker process first detects global activities; 3) then the checker process builds the ordering among global activities. Notations used in the design of OGA are listed in Table \ref{T:Notations-Model} and \ref{T:Notations-Design}.

\begin{table}
\caption{Notations in design of OGA}
\label{T:Notations-Design}
\centering
\begin{tabular}{r | p{1.6in}}\hline

Notation & Explanation \\ \hline \hline

$CurIntv$ & interval of local activity on the non-checker process \\
$flagMsgAct$ & boolean value used to denote whether there have been new message activities \\
$VC^{(k,t)}$ & vector clock timestamp on $P^{(k,t)}$ \\
$Que_{(k, t)}$ & queue for $P^{(k, t)}$ in $GA_k$ on the checker process \\
$QueLo_k, QueHi_k$ & queues for recording results of detecting $GA_k$ \\
$CurQueLo, CurQueHi$ & current global activity to be ordered \\
$PreQueLo, PreQueHi$ & previous global activity which has been ordered \\

\hline

\end{tabular}
\end{table}

\subsection{Message Activities on Non-checker Process $P^{(k,t)}$ in $GA_k$}

On the non-checker process $P^{(k,t)}$, different message activities are specified upon the beginning and ending of the local activity:
\begin{itemize}
  \item Upon $LA^{(k,t)} \uparrow$, a control messages is sent to every $P^{(k,s)} (1\leq s\leq size(GA_k), s\neq t)$, i.e., all other non-checker processes in the same global activity with $P^{(k,t)}$. The message activity here aims at building the happen-before relationship required in Equation (\ref{E:OverlappingIntervals}), in order to detect $GA_k$.
  \item Upon $LA^{(k,t)} \downarrow$, a control message is sent among every other non-checker processes $P_i (P_i \neq P^{(k,t)})$. The message activity here aims at the ordering among different global activities, as required in Equation (\ref{E:Ordering-2GA}). Meanwhile, a checking message is sent to $P_{che}$. This checking message sends vector clock timestamps $([lo, hi])$ of $I(LA^{(k,t)})$ to $P_{che}$ for the detection and ordering of global activities, as discussed in Section III.B and III.C respectively.
\end{itemize}

\noindent Boolean variable $flagMsgAct$ is used to reduce redundant message passing, as in \cite{Garg96, Huang09}. The initial value of $flagMsgAct$ is true. Pseudo codes of OGA on the non-checker process side are listed in Algorithm \ref{A:Nonchecker-Process}.

\begin{algorithm}[t]
\caption{OGA on $P^{(k, t)}$ in $GA_k$}
\label{A:Nonchecker-Process}
\begin{algorithmic}[1]

\STATE \textbf{Upon} $LA^{(k, t)}\uparrow$
\STATE $send_{control}(VC^{(k, t)})$ to each $P^{(k,s)}$ in $GA_k$, $s \neq t$;
\IF{$flagMsgAct$}
  \STATE $CurIntv.lo := VC^{(k, t)}$;
\ENDIF \\ \


\STATE \textbf{Upon} $LA^{(k, t)}\downarrow$
\STATE $send_{control}(VC^{(k, t)})$ to each $P_i (1 \leq i \leq n, P_i \neq P^{(k, t)})$;
\IF{$flagMsgAct$}
  \STATE $CurIntv.hi := VC^{(k, t)}$;
  \STATE $send_{checking}(CurIntv[lo, hi])$ to $P_{che}$;
  \STATE $flagMsgAct := false$;
\ENDIF \\ \


\STATE \textbf{Upon} $receive\_control\_msg(VC_i)$ from $P_i$
\FOR{j = 1 to n}
  \STATE $VC^{(k, t)}[j] = max\{VC^{(k, t)}[j], VC_i[j]\}$;
\ENDFOR
\STATE $flagMsgAct := true$;

\end{algorithmic}
\end{algorithm}

\subsection{Detecting Global Activities}

\subsubsection{Checking the concurrency}

Checking messages from all the non-checker processes are grouped according to the global activity they belong to. For given global activity $GA_k$, we check the concurrency among local activities based on Formula (\ref{E:OverlappingIntervals}). The checker process has a separate queue $Que_{(k,t)}$ for each $P^{(k,t)}$ in $GA_k$. Incoming checking messages are enqueued in the appropriate queue.

We assume that $P_{che}$ receives messages from each non-checker process in FIFO order as in \cite{Garg94, Garg96}. Note that this is not a restrictive assumption. We do not require FIFO for the underlying communication. $P_{che}$ needs to implement the FIFO property for efficiency purposes. If the underlying communication is not FIFO, $P_{che}$ ensures this property by using sequence numbers in messages.

Each element of $Que_{(k,t)}$ is timestamp $[lo, hi]$ of an interval. The $lo$s and $hi$s are compared to check the concurrency among intervals. The checker process reduces the number of comparisons by deleting any interval at the head of any queue whose $hi$ is not greater than $lo$ of the interval at the head of all other queues. $P_{che}$ detects $GA_k$ if it finds a set of intervals at the head of queues such that they are pairwise overlapping. The detection of concurrency is mainly based on the {\it strong conjunctive predicate} algorithm in \cite{Garg96} and our previous work \cite{Huang09}.

\subsubsection{Calculating the interval of $GA_k$}

After the detection of $GA_k$, we need to calculate $I(GA_k)$, the interval of this global activity, for further ordering of global activities. For and-activities, we need to calculate the intersection of intervals, while for or-activities, we need to calculate the union of intervals, as shown in Fig. \ref{F:Interval-Operations}.

\begin{figure}
\begin{center}
  \includegraphics[width=3in]{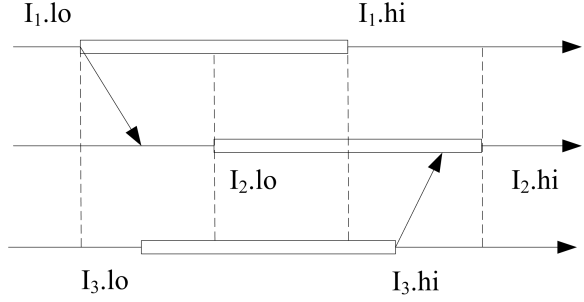}
  \caption{Calculating the interval of a global activity}
  \label{F:Calculating-GA-Interval}
\end{center}
\end{figure}

In the ideal case, for an and-activity, we calculate the latest $lo$ (every other $lo$ happens before it) and the earliest $hi$ (happening before every other $hi$). However, we may not alway be able to obtain the latest/earliest $lo$/$hi$ in asynchronous environments. For example in Fig. \ref{F:Calculating-GA-Interval}, we cannot decide which one is later for $I_2.lo$ and $I_3.lo$. Neither can we decide which one is earlier for $I_1.hi$ and $I_3.hi$. Thus, for all the $lo$s, we prune those which happens before any other $lo$ (must not be the latest), and keep all the remaining (concurrent) $lo$s. Similarly, for all the $hi$s, we prune those which ``happens after" any other $hi$ (must not be the earliest), and keep all the remaining (concurrent) $hi$s. For example in Fig. \ref{F:Calculating-GA-Interval}, we need to keep $I_2.lo$ and $I_3.lo$, as well as $I_1.hi$ and $I_3.hi$.

The or-activity is the dual of and-activity. Similar duality remains in calculating the interval of and- and or-activities. For an or-activity, we need the earliest $lo$ and the latest $hi$.

Pseudo codes for the detection of global activities are listed in Algorithm \ref{A:Detecting-GA}.

\begin{algorithm}[t]
\caption{Detecting $GA_k$ in OGA}
\label{A:Detecting-GA}
\begin{algorithmic}[1]

\STATE \textbf{Upon} receiving $CurIntv[lo, hi]$ from $P^{(k, t)}$
\STATE insert $CurIntv[lo, hi]$ to $Que_{(k, t)}$;
\IF{$CurIntv[lo, hi] \neq Que_{(k,t)}.head()$}
  \STATE return;
\ENDIF \\ /* if $CurIntv$ is the head element in $Que_{(k, t)}$, continue the checking */

\STATE $changed := \{P^{(k, t)}\}$;

\WHILE{$changed \neq \phi$}
  \STATE $newchanged := \phi$;
  \FOR{ each $P^{(k,i)}$ in $changed$ and $P^{(k, j)}$ in $GA_k$}
    \IF{$Que_{(k, j)}.head().lo \not\rightarrow Que_{(k, i)}.head().hi$}
      \STATE $newchanged := newchanged \cup \{P^{(k, i)}\}$;
    \ENDIF
    \IF{$Que_{(k, i)}.head().lo \not\rightarrow Que_{(k, j)}.head().hi$}
      \STATE $newchanged := newchanged \cup \{P^{(k, j)}\}$;
    \ENDIF
  \ENDFOR
  \STATE $changed := newchanged$;
  \FOR{each $P^{(k, i)}$ in $changed$}
    \STATE $delete\_head(Que_{(k, i)})$;
  \ENDFOR
\ENDWHILE \\ /* if $GA_k$ is detected */

\IF{$\forall i$, $Que_{(k, i)}$ is not empty}
  \STATE calculate $I(GA_k)$;
  \STATE enqueue each $lo$ and $hi$ remained after the pruning to $QueLo(GA_k)$ and $QueHi(GA_k)$ respectively;
\ENDIF

\end{algorithmic}
\end{algorithm}

\subsection{Ordering Global Activities}

The essential issue in ordering two global activities is to establish the relative order between $I(GA_k).hi$ and $I(GA_{k+1}).lo$. As discussed in the previous section, we may encounter multiple (concurrent) $lo$s and $hi$s when detecting global activities. We have stored all these $lo$s and $hi$s in appropriate queues as shown in Algorithm \ref{A:Detecting-GA}. Now, we compare all the stored $lo$s and $hi$s. This comparison continues until $I(GA_k).hi \rightarrow I(GA_{k+1}).lo$ is established for every stored $hi$ and $lo$. When we reach the last global activity, we finish the ordering of global activities. Pseudo codes for the ordering of global activities are listed in Algorithm \ref{A:Ordering-GA}.

\begin{algorithm} [t]
\caption{Ordering global activities in OGA}
\label{A:Ordering-GA}
\begin{algorithmic}[1]

\WHILE{$index \leq m$}
  \REPEAT
    \STATE $get\_gloabl\_activity(index)$ and copy the results in $QueLo$ and $QueHi$ to $CurQueLo$ and $CurQueHi$ respectively;
  \UNTIL{$\forall \ VC_{pre} \in PreQueHi, \forall \ VC_{cur} \in CurQueLo$, $VC_{pre} \leq VC_{cur}$;}

  \STATE $PreQueLo := CurQueLo$;
  \STATE $PreQueHi := CurQueHi$;

  \STATE $++index$;
\ENDWHILE

\end{algorithmic}
\end{algorithm}

\subsection{Discussions}

The number of comparisons for detecting a global activity is $O(s^2p)$, where $s$ is the upper bound of size of the global activity, $p$ is the upper bound of length for each queue in detecting the global activity.
The number of comparisons for ordering global activities is $O(s^2m)$. On the normal process side, the number of message activities is $O(p)$. Note that existing work may impose less message complexity, but they rely on the assumption of a global clock or synchronized interactions. The message complexity of OGA is mainly due to building the happen-before relationship between $lo$s and $hi$s, which is a requisite for detecting temporal properties in asynchronous environments.

We assumed reliability of message passing. Note that even with this assumption, we cannot guarantee correct ordering of global activities. Without this assumption, we only need to revise our algorithm to tolerate incomplete message information. Rationale of our algorithm remains the same. The probability of detecting global activities is analyzed in our previous work \cite{Huang09}. In Section V, we further evaluate OGA by experiments.

\section{Implementation}

\begin{figure}[t]
\begin{center}
  \includegraphics[width=3in]{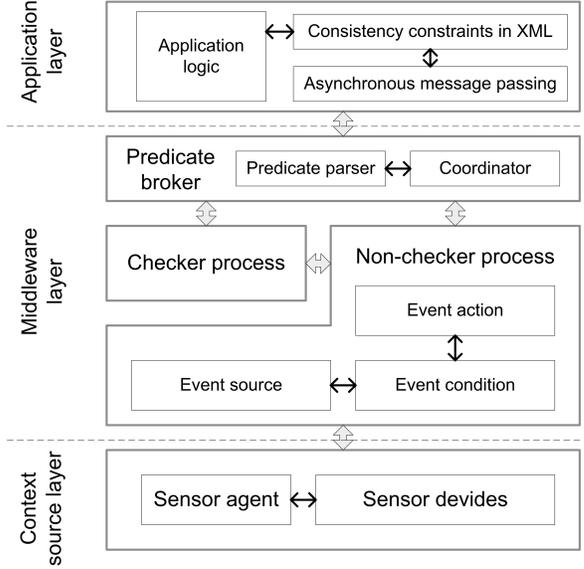}
  \caption{System architecture of MIPA}
  \label{F:MIPA}
\end{center}
\end{figure}

The OGA algorithm we propose assumes the availability of an underlying middleware infrastructure for asynchronous consistency checking of pervasive context. We have developed such a middleware named Middleware Infrastructure for Predicate detection in Asynchronous environments (MIPA) \cite{MIPA}. From MIPA's point of view, a pervasive computing environment is composed of an {\it application layer}, a {\it middleware layer} and a {\it context source layer}, as shown in Fig. \ref{F:MIPA}.

The middleware layer is the kernel part of MIPA. Its fundamental functionalities include:
\begin{itemize}
  \item {\it Predicate broker}. The predicate broker accepts consistency constraints specified by the context-aware application. It first parses the consistency constraint, and then initiates the non-checker processes and the checker process accordingly.
  \item {\it Non-checker process}. The non-checker process monitors the local predicate value based on the Event-Condition-Action (ECA) mechanism \cite{Chan03}. It accepts source contextual events from the corresponding sensor agents. The local predicate serves as the event condition. When value of the local predicate changes, the consistency checking algorithm on the checker process side is triggered. The non-checker process sends messages to build the requisite happen-before relationship. It also sends checking message to the checker process, which finally decides whether the consistency constraint is satisfied.
  \item {\it Checker process}. The checker process collects vector clock timestamps of local contextual activities. It executes the predicate detection algorithm to decide whether the application-specified consistency constraint is satisfied. The checking result is sent back to the application via the predicate broker.
\end{itemize}

We implement the OGA algorithm on MIPA, and conduct the experimental evaluation, as discussed in detail in the next section.

\section{Experimental Evaluation}

In the previous Section III, we presented design of the OGA algorithm. However, does OGA work in pervasive computing environments? Specifically, can OGA achieve accurate ordering of global activities? We investigate these issues by experiments in this section.

\subsection{Experiment setup}

The experimental evaluation is based on a smart-lock scenario first investigated in our previous work \cite{Wu08}. In this scenario, a smart-lock application automatically locks the office when the users leaves, i.e., when user's location changes from `office' to `corridor'. To deal with noisy sensor readings, the user's location context is detected by both an RFID reader \cite{Mamei07, Floerkemeier07} and a light sensor. User's location is detected by the global activity ``$GA_1$ = (the user is detected by the RFID reader in the office) $\wedge$ (the user is detected by the light sensor in the office)" and ``$GA_2$ = (the user is detected by the RFID reader in the corridor) $\wedge$ (the user is detected by the light sensor in the corridor)". User's behavior of leaving the office is delineated by the behavioral consistency constraint ``$GA_1\prec GA_2$".

We model user's stay in the office based on the queueing theory \cite{Bunday96}. Specifically, a queue of intervals with Poisson arrival rate $\frac{1}{600s}$ is adopted. The duration of intervals follows the exponential distribution of rate $\frac{1}{300s}$. We model the message delay by the exponential distribution. Note that the distribution of message delay is affected by implementation of the underlying network layers (e.g., the MAC or routing layer), and greatly varies in different scenarios. Though it is doubted whether there exists a universal model of message delay, the exponential distribution is widely used and evaluated by both simulations and experiments \cite{Duffield00,Duffield01}. Our experiment methodology is also applicable when the message delay follows other types of distributions.

In the evaluation, we study how asynchrony of the computing environment affects the performance of OGA. The update interval of sensor data dissemination and the message delay are varied. This issue is critical since the asynchrony is the primary motivation of our work. We also study the effect of tuning the duration of the user's stay in the office. This duration decides how frequently the user leaves the office.

Performance of the OGA algorithm is measured by the probability of correct ordering of global activities in asynchronous environments. We obtain this probability of correct ordering by calculating the ratio of $\frac{Num_{OGA}}{Num_{phy}}$. Here, $Num_{OGA}$ denotes how many times OGA detects the ordering of global activities. $Num_{phy}$ denotes the number of the ordering of global activities, obtained from physical time of each local contextual activity. Detailed experiment configurations are listed in Table \ref{T:Exp}.

\begin{table}[t]
\begin{center}
\caption{Experiment configurations}
\label{T:Exp}
\begin{tabular} {r|l}
\hline
Parameter                           & Value \\
\hline
\hline
Number of global activities & 2 \\
Number of non-checker processes & 4 \\
Lifetime of application & $20\times 24$ h \\
Average stay in office & $600 s$ \\
Average stay out of office  & $300 s$ \\
Update interval of sensors & $1 s \sim 5400 s$ \\
Average message delay & $0.06 s \sim 300 s$\\
\hline
\end{tabular}
\end{center}
\end{table}

\subsection{Effects of Tuning the Update Interval}

\begin{figure}
\begin{center}
  \includegraphics[width=3in]{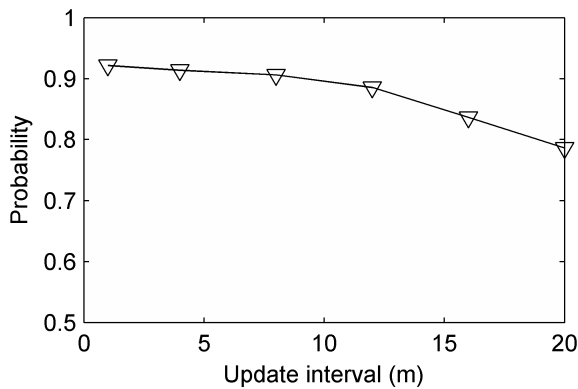}
  \caption{Update interval (0m $\sim$ 20m)}
  \label{F:Update-Interval-Small}
\end{center}
\end{figure}

\begin{figure}
\begin{center}
  \includegraphics[width=3in]{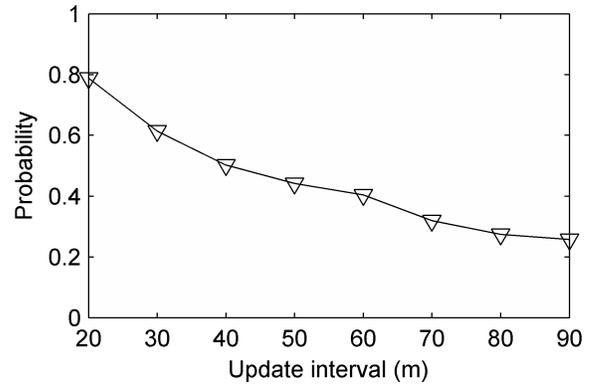}
  \caption{Update interval (20m $\sim$ 90m)}
  \label{F:Update-Interval-large}
\end{center}
\end{figure}

In this experiment, we study the effect of tuning length of the update interval of the sensors. We find that the increase in the update interval results in monotonic decrease in the probability of correct ordering of global activities, as shown in Fig. \ref{F:Update-Interval-Small} and \ref{F:Update-Interval-large}. The is mainly because the increase of update interval adds to the asynchrony of the environment. Specifically, the probability of correct ordering is high (over 90\%) when the update interval is less than 10 minutes, as shown in Fig. \ref{F:Update-Interval-Small}. When the update interval gets longer than the average duration of the user's stay in the office (10 minutes), the probability begins to decrease much more quickly, as shown in Fig. \ref{F:Update-Interval-Small}. When we increase the update interval to a large value (up to 90 minutes), the probability may decrease to around 20\%, as shown in Fig. \ref{F:Update-Interval-large}.

In summary, the evaluation results here show the impact of asynchrony of the environment on the checking of behavioral consistency constraints. They also show that OGA can achieve accurate checking, even when the update interval is reasonably long.

\subsection{Effects of Tuning the Message Delay}

\begin{figure}
\begin{center}
  \includegraphics[width=3in]{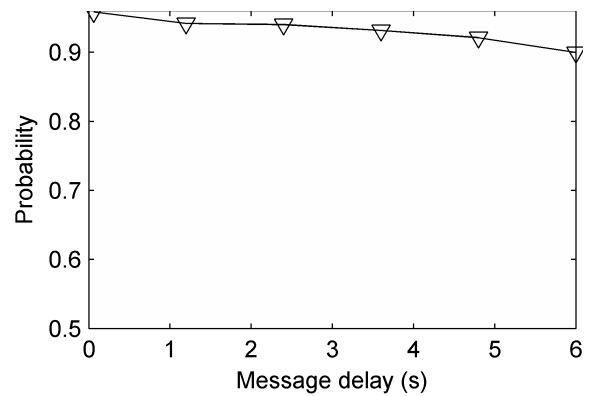}
  \caption{Message delay (0s $\sim$ 6s)}
  \label{F:Msg-Delay-Small}
\end{center}
\end{figure}

\begin{figure}
\begin{center}
  \includegraphics[width=3in]{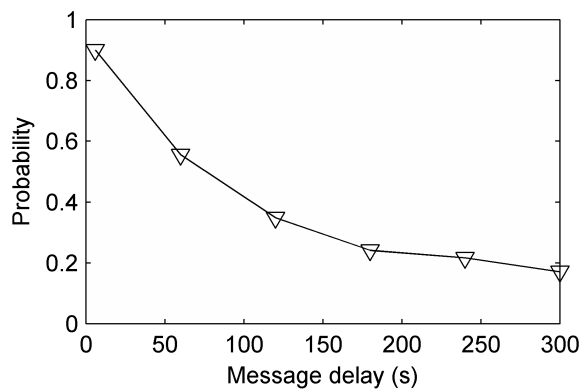}
  \caption{Message delay (6s $\sim$ 300s)}
  \label{F:Msg-Delay-Large}
\end{center}
\end{figure}

In this experiment, we study how the message delay affects the performance of OGA. We find that when encountered with reasonably long message delay (less than 1s), the probability of correct detection is quite high (a little less than 100\%), as shown in Fig. \ref{F:Msg-Delay-Small}. Only when the delay goes up to more than 1 minute, the probability begins to significantly decrease, as shown in Fig. \ref{F:Msg-Delay-Large}. Note that though the message delay usually does not go up to several minutes, we increase the message delay to large values here to explore its impact on the performance of OGA.

Combining the results in Fig. \ref{F:Msg-Delay-Small} and \ref{F:Msg-Delay-Large}, we also find that the message delay results in monotonic decrease in the probability of correct ordering of global activities, mainly due to the increase in the asynchrony of the environment. However, the impact of the message delay is comparatively less than that of the update interval.

\subsection{Effects of tuning the Duration of User's Stay in the Office}

\begin{figure}
\begin{center}
  \includegraphics[width=3in]{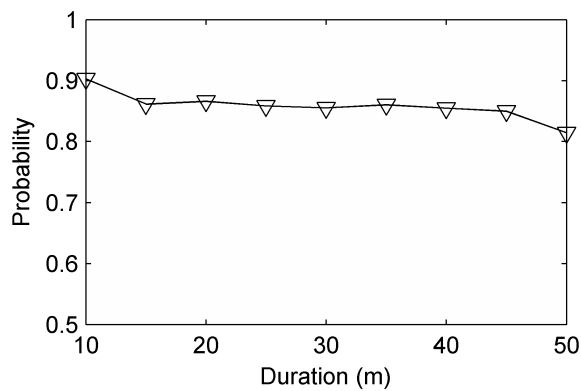}
  \caption{Duration of user's stay in the office}
  \label{F:Stay-Interval}
\end{center}
\end{figure}

In this experiment, we tune the duration of user's stay in the office. We find that tuning the duration does not has as much impact as that of tuning the update interval and the message delay, as shown in Fig. \ref{F:Stay-Interval}. The probability of correct detection slowly decreases as the duration increases. The probability first decreases as the duration increases to 15 minutes. Then it remains relatively stable. The probability decreases again when the duration increases to 50 minutes.

The duration of stay does not affect the asynchrony of the environment, thus imposing less impact on the performance of OGA. The probability of correct detection slightly decreases mainly because when the duration of stay increases, the user leaves the office less frequently. The number of global activities which can be ordered by OGA decreases.

\subsection{Lessons Learned}

Based on the experimental evaluation, we first show the impact of asynchrony in the pervasive computing environment on context consistency checking, which justifies the basic motivation of our work. We also demonstrate the performance of OGA in pervasive scenarios. Specifically, OGA achieves high probability of ordering global activities, even when faced with reasonably long update interval and message delay, as well as when faced with different frequencies of contextual behaviors.

\section{Related Work}

Many existing studies on context-aware computing are concerned with middleware infrastructures that support collection and management of contexts \cite{Murphy06, Costa07, Bettini08, Lehmann04, Grossmann05, Roy06, Li06}. Various schemes have been proposed for context consistency checking over context-aware middleware infrastructures. In \cite{Xu05}, consistency constraints were modeled by tuples, and consistency checking was based on comparison among elements in the tuples. In \cite{Xu06}, consistency constraints were expressed in first-order logic, and an incremental consistency checking algorithm was proposed. In \cite{Huang08}, a probabilistic approach is proposed to further improve the effectiveness of consistency checking. In \cite{Bu06, Bu06a}, consistency constraints were expressed by assertions. However, existing schemes do not sufficiently consider the temporal relationships among the contexts. It is implicitly assumed that the contexts being checked belong to the same snapshot of time. Such limitations make these schemes do not work in asynchronous pervasive computing environments \cite{Sama08, Huang09, Kaveti09}.

In asynchronous environments, the concept of temporal ordering of events must be carefully reexamined \cite{Lamport78}. The happen-before relationship intrinsic in message passing is a promising solution to context consistency checking in asynchronous pervasive computing environments. In our previous work \cite{Huang09}, the {\it Concurrent Event Detection for Asynchronous consistency checking} (CEDA) algorithm was proposed to detect concurrent contextual activities in asynchronous pervasive computing environments. CEDA explicitly checks whether contexts being checked belong to the same snapshot of time based on the happen-before relationship among the beginning and ending of contextual activities. However, behavior patterns of contexts cannot be specified and checked by CEDA. In this paper, we study how to check behavioral patterns of contexts based on the ordering of global contextual activities.

\section{Conclusion and Future Work}

In this paper, we study how to check the behavior patterns of contexts in asynchronous pervasive computing environments. Toward this objective, our contribution is three-fold: 1) we delineate behavioral patterns of contexts by the ordering of global activities; 2) we propose the OGA algorithm to check behavioral constraints of context consistency in asynchronous pervasive computing environments; 3) we implement the MIPA middleware infrastructure for asynchronous consistency checking of pervasive context. The OGA algorithm is developed and evaluated over MIPA.

In our future work, we will study the design of a general framework, covering various existing predicates, as well as their checking algorithms. The framework will help us better understand the pervasive computing environment from a predicate detection perspective. We will also extend our middleware infrastructure MIPA to support the proposed framework.

\section*{Acknowledgment}

The authors would like to thank Dr. Chang Xu, Dr. Hung Keng Pung and Dr. S.C. Cheung for their constructive comments and suggestions.

This work is supported by the National 973 Program of China (2009CB320702), the National Natural Science Foundation of China (No. 60903024, 60736015) and the ``Climbing" Program of Jiangsu Province, China (BK2008017).

%
\bibliographystyle{IEEEtran}
\bibliography{TR090709}





\end{document}